\documentclass[10pt, journal]{IEEEtran}
\IEEEoverridecommandlockouts
\usepackage{amsmath}
\usepackage{amssymb}
\usepackage{url}
\usepackage{cite}
\usepackage{flushend}
\usepackage{makecell}
\usepackage{subfigure}
\usepackage{amsfonts,amsthm}

\usepackage{algorithmic}
\usepackage{algorithm}

\usepackage{graphicx}
\usepackage{textcomp}
\usepackage{float}
\usepackage{booktabs}
\usepackage{bm}
\usepackage{float}  %设置图片浮动位置的宏包
\usepackage{pifont} % 添加在已加载的宏包之后（如 fontenc 下方）
\usepackage{wasysym}

\usepackage{afterpage} 
\usepackage{multirow}
\usepackage[normalem]{ulem}
\useunder{\uline}{\ul}{}
\usepackage{xcolor}
\def\BibTeX{{\rm B\kern-.05em{\sc i\kern-.025em b}\kern-.08em
    T\kern-.1667em\lower.7ex\hbox{E}\kern-.125emX}}
\allowdisplaybreaks

\usepackage{tablefootnote}

\begin{document}

\title{Contextual Memory-Enhanced Source Coding for Low-SNR Communications}

\author{Ziqiong Wang, Rongpeng Li, Zhifeng Zhao, and Honggang Zhang
    \thanks{Z. Wang and R. Li are with the College of Information Science and Electronic Engineering, Zhejiang University (email: \{wangziqiong, lirongpeng\}@zju.edu.cn). Z. Zhao is with Zhejiang Lab as well as the College of Information Science and Electronic Engineering, Zhejiang University (email: zhaozf@zhejianglab.org). H. Zhang is with the Macau University of Science and Technology, China (email:  honggang.zhang@ieee.org).}
}

\maketitle
   
\begin{abstract}
% Separate Source–Channel Coding (SSCC) remains attractive for text transmission due to its modularity and compatibility with mature entropy coders and powerful channel codes. Nevertheless, SSCC often suffers from a pronounced cliff effect in low Signal-to-Noise Ratio (SNR) regimes, where residual bit errors after channel decoding can catastrophically break lossless source decoding, especially for Arithmetic Coding (AC) driven by Large Language Models (LLMs). 
Separate Source-Channel Coding (SSCC) remains vulnerable in noisy text transmission due to the fragility of autoregressive source decoding, especially when Arithmetic Coding (AC) relies on Large Language Model (LLM)-based probability estimation. 
% Existing remedies either strengthen channel decoding based solely on channel observations or introduce contextual information only at the receiver for post-hoc correction, yet neither fully addresses the fragility of source probability modeling under residual channel errors.
This letter proposes a Memory-Augmented Source Coding (MASC) scheme that internalizes contextual patterns into a source model. Specifically, MASC employs a shared Parameterized Contextual Memory (PCM) for multi-order $n$-gram patterns, and a Mixture-of-Memory-Experts Router (MMER) for sparse, hidden-state-dependent routing over memory experts. This adaptive activation refines source probability estimation, shortens codelength, and mitigates decoding sensitivity to residual channel errors. Experiments over Rayleigh fading and AWGN channels demonstrate its effectiveness. 
\end{abstract}


\section{Introduction}\label{sec1_Introduction}

\IEEEPARstart{S}{emantic} Communications (SemCom) have emerged as a transformative paradigm~\cite{luRethinkingModernCommunication2023,semantics2} that leverages Joint Source-Channel Coding (JSCC) to combat channel non-stationarity~\cite{JSCC3}. Nevertheless, Separate Source-Channel Coding (SSCC) remains a cornerstone of practical communication systems due to its inherent modularity and seamless integration with existing industrial standards~\cite{SSCC1}. %By leveraging mature neural compressors~\cite{LLM_coding2,LLM_coding3,LLM_coding4,LLM_coding5} and capacity-approaching channel codes~\cite{channel_code}, SSCC can theoretically achieve near-error-free, bit-exact reconstruction in high Signal-to-Noise Ratio (SNR) regimes. 
However, the primary obstacle to deploying neural SSCC in practical noisy environments is the ``cliff effect''. Due to the sequential dependency of compressed bitstreams, even a small number of residual bit errors after channel decoding may catastrophically break subsequent source decoding. Conventional remedies primarily focus on the channel-coding side, employing advanced decoders such as the Error Correction Code Transformer (ECCT)~\cite{ECCT} to mitigate noise through code-aware inductive biases. Nevertheless, such decoders remain fundamentally constrained by channel observations alone~\cite{ECCT_limit}, without introducing explicit external information.

To bridge this gap, recent advances have leveraged contextual information~\cite{context1} at the receiver as a backward-compatible means to enhance robustness. Representative techniques include candidate maintenance~\cite{candidate_decoding}, backtracking~\cite{backtracking}, and stochastic sampling~\cite{sampling}. Along these lines, our recent study further proposed a receiver-side in-context decoding framework that couples ECCT-assisted channel decoding with Large Language Model (LLM)-based source decoding, where contextual information and channel-decoding reliability are jointly used to enhance performance~\cite{ICD}. However, its effectiveness still hinges on sufficiently reliable contextual information at the receiver, which may be difficult to guarantee in practice. More fundamentally, such context is introduced merely as external side information at the receiver, rather than internalized into a shared memory that directly supports source probability modeling at both the transmitter and the receiver.

% These limitations suggest that the key issue is not the absence of contextual information, but the manner in which it is incorporated into the communication system. In many semantic communication frameworks, context or knowledge is introduced through external knowledge bases and retrieval mechanisms to support semantic inference or receiver-side recovery. However, this paradigm is not fully aligned with the present problem. Rather than injecting external knowledge to compensate for missing semantics after transmission. rather, our objective is to improve the probability model that governs source coding and autoregressive source decoding. This requires contextual regularities to be internalized into the source model and shared consistently by both ends of the transceiver.

% In this paper, we propose a Memory-Augmented Source Coding (MASC) scheme for SSCC-based transmission to alleviate the cliff effect. Specifically, MASC is built upon a Transformer backbone and a shared Parameterized Contextual Memory (PCM). At each coding step, Mixture-of-Memory-Experts Router (MMER) within PCM leverages the hidden state from the previous Transformer block to activate a sparse subset of relevant memory experts and retrieves multi-order $n$-gram memory representations of recurring local semantic patterns. The retrieved memories are then fused with the hidden state and injected into the subsequent Transformer block for source modeling. In comparison to existing works in the literature, the contribution of this paper can be summarized as follows.
In this letter, we propose a Memory-Augmented Source Coding (MASC) framework, which leverages a shared Parameterized Contextual Memory (PCM) to alleviate the cliff effect. Departing from conventional Engram-based architectures~\cite{Engram} that rely on static look-up tables, MASC introduces a Mixture-of-Memory-Experts Router (MMER) within the PCM architecture. Rather than performing an exhaustive search across all stored patterns, the MMER leverages the latent hidden states of the previous Transformer block to adaptively and sparsely activate a subset of memory experts relevant to the current coding step. This mechanism enables MASC to retrieve multi-order $n$-gram representations that encompass a wide spectrum of semantic granularity. By injecting the retrieved memories into the subsequent Transformer block, MASC transforms transient context into persistent and trainable representations. Consequently, the shared memory provides a stable semantic foundation for probability modeling that effectively shortens the average codelength and thereby mitigates the fragility of the transmission against channel noise. We conduct extensive evaluations and demonstrate the effectiveness and superiority of MASC over conventional SSCC baselines, represented by Huffman-SSCC and ECCT-aided scheme~\cite{SSCC1}, typical JSCC schemes, including DeepSC~\cite{DeepSC}, Universal Transformer (UT)~\cite{UT}, and UT with quantization~\cite{UT_quanti}, as well as receiver-side in-context decoding methods such as ICD~\cite{ICD}.

% \begin{itemize}
%     \item We introduce a MASC scheme for robust SSCC-based transmission. Different from receiver-side contextual or knowledge-assisted correction methods, MASC enhances robustness by internalizing context into the probability modeling process shared by the transceiver.
%     \item We develop an MMER module for sparse adaptive routing over multi-order $n$-gram memories. By selecting relevant experts according to the hidden states, MMER captures multi-scale semantic patterns with high computational efficiency.
%     \item We conduct extensive evaluations and demonstrate the effectiveness of MASC over conventional SSCC baselines, represented by Huffman-SSCC and ECCT-aided scheme~\cite{SSCC1}, representative JSCC schemes, including DeepSC~\cite{DeepSC}, Universal Transformer (UT)~\cite{UT}, and UT with quantization~\cite{UT_quanti}, as well as receiver-side in-context decoding methods such as ICD~\cite{ICD}.
% \end{itemize}

The remainder of the letter is organized as follows. Sec. \ref{sec2_System Model} briefly introduces the system model and formulates the problem. Sec. \ref{sec3_Framework} presents the overview of our proposed MASC scheme. In Sec. \ref{sec4_Experiment}, we elaborate on the experimental results and discussions. Finally, Sec. \ref{sec5_Conclusions} concludes the letter.

\begin{figure}[!tb]
\centering
\includegraphics[width=\linewidth]{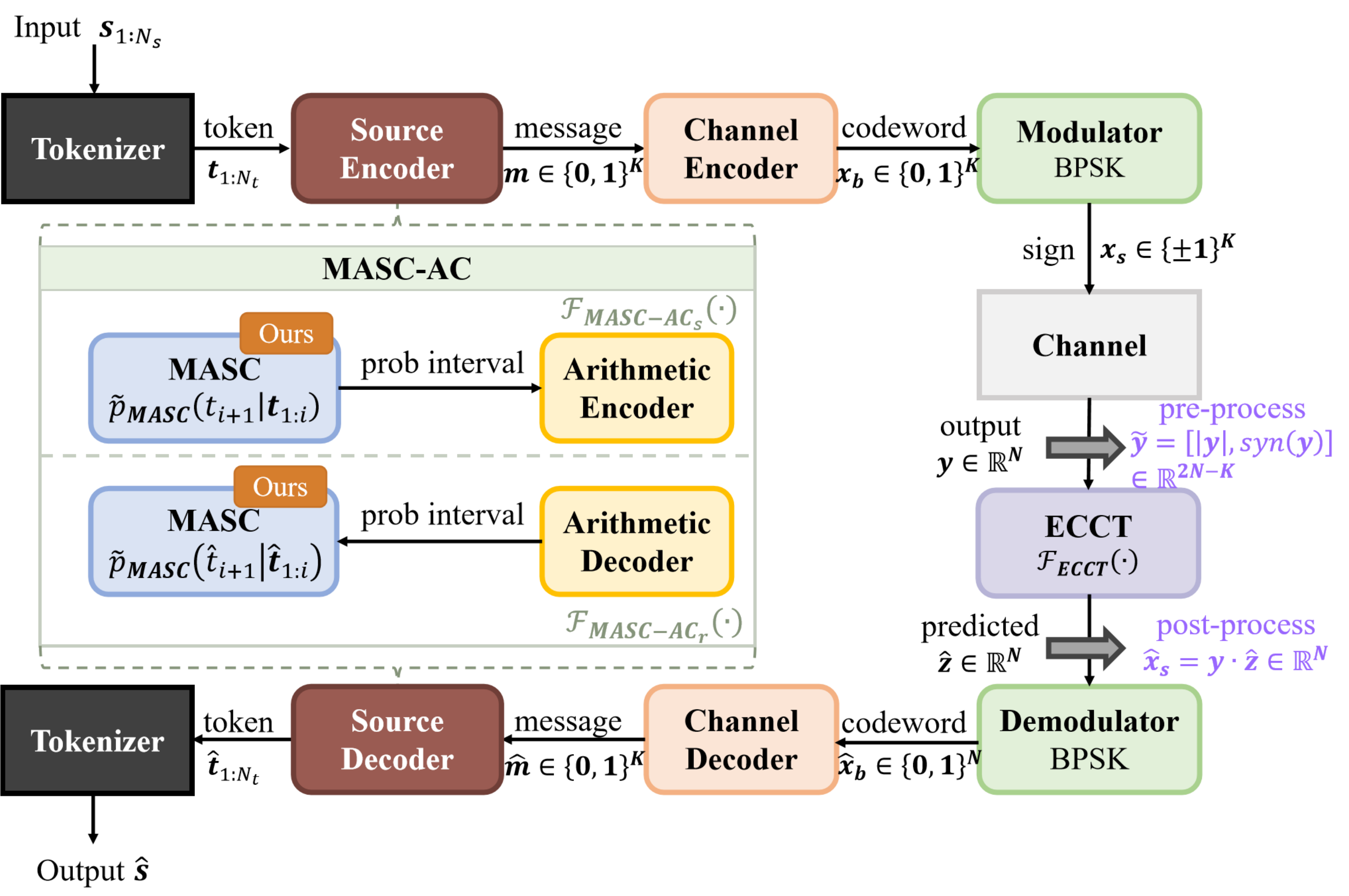}
\caption{Framework of the Proposed MASC-Based SSCC System.}
\label{fig:system model}
\end{figure}
\section{System Model and Problem Formulation}\label{sec2_System Model}

\subsection{System Model}

% We consider an SSCC-based text transmission framework as illustrated in Fig.~\ref{fig:system model}. For the input text sequence $\textbf{s}_{1:N_s}$, a lossless source encoder compresses it into a binary message
% \begin{equation}
% \textbf{m}=\mathcal{F}_{\mathrm{LLM-AC}_s}(\textbf{s}_{1:N_s})\in\{0,1\}^{K},
% \end{equation}
% where as mentioned in Sec. \ref{sec2:LLM_coding}, $\mathcal{F}_{\mathrm{LLM-AC}_s}(\cdot)$ denotes a LLM-compatible source encoder that maps text into a length-$K$ bitstream through tokenization and probability-driven AC. 
% The same pipeline can also be applied to other discrete sources (e.g., images) after appropriate lossless formatting or compression.
We consider an SSCC-based text transmission framework as illustrated in Fig.~\ref{fig:system model}. For the $N_s$-length input text sequence $\textbf{s}_{1:N_s}$, model-based Arithmetic Coding (AC) is employed for source coding~\cite{SSCC1}. Specifically, given a tokenizer vocabulary $\mathcal{D}=\{D_1,\ldots,D_\tau\}$, the source sequence is first tokenized into $\textbf{t}_{1:N_t}$ with $t_i\in\mathcal{D}$, and a pre-trained LLM provides the conditional probabilities $\tilde{p}_{\mathrm{LLM}}(t_{i+1} \mid \textbf{t}_{1:i})$ that drive the AC procedure. The overall source encoding process can be abstracted as
\begin{equation}
\textbf{m}=\mathcal{F}_{\mathrm{LLM-AC}_s}(\textbf{s}_{1:N_s})\in\{0,1\}^{K},
\end{equation}
where $\mathcal{F}_{\mathrm{LLM-AC}_s}(\cdot)$ denotes an LLM-compatible source encoder that maps text into a length-$K$ bitstream through tokenization and probability-driven AC. 
The same pipeline can also be applied to other discrete sources (e.g., images) after appropriate lossless formatting or compression.

The message $\textbf{m}$ is then protected by an $(N,K)$ Low-Density Parity-Check (LDPC) channel code as
\begin{equation}
\textbf{x}_{b}=\textbf{m}\mathbf{G}\ (\mathrm{mod}\ 2)\in\{0,1\}^{N},
\end{equation}
where $\mathbf{G}\in\{0,1\}^{K\times N}$ is the generator matrix. The corresponding parity-check matrix $\mathbf{H}\in\{0,1\}^{(N-K)\times N}$ satisfies
\begin{equation}
\mathbf{G}\mathbf{H}^{\top}=\mathbf{0}.
\end{equation}
Notably, other error correction codes such as Polar codes can be applied as well \cite{ECCT}. 
% The binary codeword $\textbf{x}_{b}$ is modulated via BPSK into $\textbf{x}_s=1-2\textbf{x}_{b}\in\{\pm 1\}^{N}$ and transmitted over the channel, as defined in Eq. \eqref{eq:channel}.
The binary codeword $\textbf{x}_{b}$ is modulated via Binary Phase-Shift Keying (BPSK) as
\begin{equation}
\textbf{x}_s=1-2\textbf{x}_{b}\in\{\pm 1\}^{N}, 
\end{equation}
and transmitted over the channel according to
\begin{equation}
\label{eq:channel}
\textbf{y}=h\textbf{x}_s+\textbf{z},\qquad \textbf{z}\sim\mathcal{N}(\textbf{0},\sigma_n^2\mathbf{I}),
\end{equation}
where $h$ denotes the channel coefficient and $\textbf{z}$ is the additive Gaussian noise. 
% Following~\cite{ECCT}, the channel observation can also be expressed in a multiplicative form $\textbf{y}=\textbf{x}_s\odot\tilde{\textbf{z}}$, where $\tilde{\textbf{z}}\in\mathbb{R}^{N}$ denotes an equivalent multiplicative disturbance.

At the receiver, we adopt the Error Correction Code Transformer (ECCT)\footnote{Complete architectural and training details of ECCT can be found in~\cite{ECCT}.}~\cite{ECCT} to mitigate channel impairments. By exploiting the channel observations $\textbf{y}$ alongside the algebraic constraints of $\mathbf{H}$, ECCT produces an estimated disturbance $\hat{\textbf{z}}$ as well as a recovered BPSK-coded sequence $\hat{\textbf{x}}_s$. The binary codeword estimate is then obtained via hard-decision demodulation as
\begin{equation}
\hat{\textbf{x}}_{b}=\text{sign\_to\_bin}(\hat{\textbf{x}}_{s})\in\{0,1\}^{N},
\end{equation}
where $\text{sign\_to\_bin}(\hat{\textbf{x}}_{s}) \triangleq \frac{1}{2}\big(\textbf{1}-\text{sign}(\hat{\textbf{x}}_{s})\big)\in\{0,1\}^{N}$. 
We consider a systematic channel encoder, where the information bits occupy the first $K$ positions of the codeword.
In this case, recovering the information bitstream reduces to extracting the first $K$ bits as 
\begin{equation}
\hat{\textbf{m}}=\hat{\textbf{x}}_{b,1:K}\in\{0,1\}^{K}.
\end{equation}

Given the channel decoding result $\hat{\textbf{m}}$, an LLM-based source decoder $\mathcal{F}_{\text{LLM-AC}_r}(\cdot)$ can eventually produce a reconstructed sequence $\hat{\textbf{s}}$ as follows 
\begin{equation}
\hat{\textbf{s}}
=\mathcal{F}_{\text{LLM-AC}_r}\!\left(\hat{\textbf{m}}\right).
\end{equation}

\begin{figure*}[!tb]
    \centering
    \includegraphics[width=0.9\linewidth]{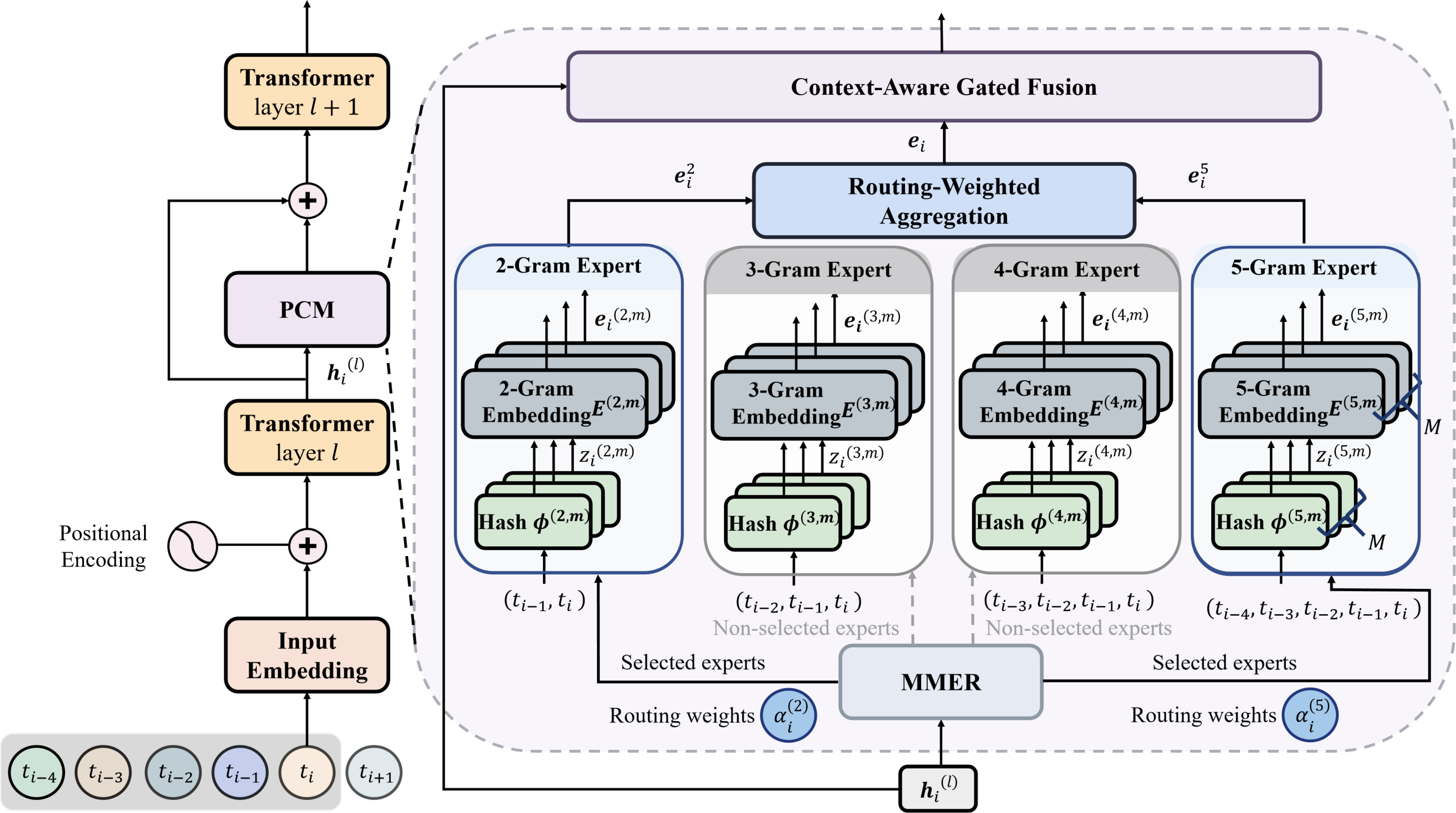}
    \vspace{-0cm}
    \caption{The MASC Architecture.}
    \vspace{-.7cm}
    \label{fig:MASC}
\end{figure*}
\subsection{Problem Formulation}

Conventional SSCC systems often suffer from a \emph{cliff effect} in low-SNR scenarios. As illustrated by the fourth purple dash-dotted curve in Fig. \ref{fig:overall_performance_rayleigh} and Fig. \ref{fig:overall_performance_awgn}, once the channel quality drops below a specific SNR threshold, a small number of residual bit errors may catastrophically disrupt the subsequent LLM-based arithmetic decoding process. The primary reason lies in the fact that an incorrect early bit can shift the LLM-based decoder to a wrong sub-interval for arithmetic decoding, which then perturbs subsequent token boundary decisions and triggers error propagation across many tokens. Consequently, enhancing the quality of the source probability model $\tilde{p}_{\mathrm{LLM}}(t_{i+1} \mid \textbf{t}_{1:i})$ is critical, as a more precise estimate yields a shorter compressed bitstream, which in turn reduces the vulnerability of the transmitted message to channel impairments. 

% Motivated by this principle, we move beyond receiver-side contextual correction and instead consider a transceiver-level memory-augmented source coding paradigm. Specifically, rather than treating contextual information as auxiliary side information available only at the receiver, we seek to internalize recurring local semantic patterns into a shared PCM that can be jointly exploited by both the transmitter-side source encoder and the receiver-side source decoder. Let PCM contain a set of multi-order $n$-gram memory components, denoted by

To address these challenges, we propose the MASC scheme. As a transceiver-level paradigm, MASC integrates a Transformer-based backbone with a specialized PCM component to refine the source probability model. Within the PCM architecture, an MMER, denoted by $\mathcal{R}(\cdot)$, is employed to adaptively manage a set of multi-order $n$-gram memory modules:
\begin{equation}
\mathcal{M}=\left\{\mathcal{M}^{(n)}\right\}_{n=2}^{5},\label{eq:backbone_memory}
\end{equation}
where $\mathcal{M}^{(n)}$ corresponds to the memory component associated with the $n$-gram order\footnote{Different from the original Engram backbone, which only uses the $2$-gram and $3$-gram memories, MASC extends the memory orders to $5$ for better capturing long range contextual patterns.}. To avoid activating all such components at every coding step, the MMER adaptively selects a sparse subset of relevant memory experts according to the current hidden state $\textbf{h}_i$:
\begin{equation}
\mathcal{A}_i=\mathcal{R}(\textbf{h}_i), 
\qquad
\mathcal{A}_i\subseteq \mathcal{M},
\end{equation}
where $\mathcal{A}_i$ denotes the activated memory subset for the \(i\)-th token.

Under this architecture, the MASC-driven source encoding and decoding are formulated as $\textbf{m}=\mathcal{F}_{\mathrm{MASC-AC}_s}\!\left(\textbf{s}_{1:N_s}\right)$ and $\hat{\textbf{s}} =\mathcal{F}_{\mathrm{MASC-AC}_r}\!\left(\hat{\textbf{m}}\right)$, respectively. Our objective is to jointly optimize the memory parameters within the PCM and the sparse routing strategy of the MMER to maximize reconstruction fidelity while adhering to a strict complexity budget. This is formulated as the following optimization problem:

% Formally, $\mathcal{F}_{\mathrm{MASE}}(\cdot)$ and $\mathcal{F}_{\mathrm{MASD}}(\cdot)$ preserve the same arithmetic source coding and decoding procedures as the generic operations $\mathcal{F}_{s}(\cdot)$ and $\mathcal{F}_{r}(\cdot)$ introduced in the system model. Their distinction lies in the conditional distribution used to drive the coding process, with the conventional formulation relying on the pretrained LLM and the proposed formulation employing a memory-augmented source model.

% To this end, we design a computationally efficient memory-augmented source coding scheme that improves the conditional distribution $\tilde{p}_{\mathrm{MASC}}(t_{i+1} \mid \textbf{t}_{1:i})$, shortens the resulting arithmetic codewords, and enhances the final reconstruction reliability under noisy channels. This corresponding objective can be expressed as

\begin{equation}
\min_{\Theta}
\ \mathbb{E}\!\left[d\!\left(\textbf{s}_{1:N_s},\hat{\textbf{s}}\right)\right]
\quad
\text{s.t.}\quad
\mathrm{Cost}(\mathcal{R}) \le \mathcal{B},
\label{eq:problem_formulation_mascc}
\end{equation}
where $\Theta$ denotes the learnable parameters of MASC, $d(\cdot,\cdot)$ measures the reconstruction mismatch between the original source and the recovered sequence, and $\mathcal{B}$ denotes the allowable routing complexity budget. 
% The concrete design of the proposed MASC scheme will be presented in Sec.~\ref{sec3_Framework}.

\section{Memory-Augmented Source Coding}\label{sec3_Framework} 

% In this section, we present the proposed MASC scheme. %Architecturally, MASC integrates a Transformer backbone with a PCM, where the Transformer models global autoregressive dependencies and PCM captures recurring local semantic patterns through multi-order memories. The overall framework of MASC is 
% As depicted in Fig.~\ref{fig:MASC}, the proposed scheme operates by selecting the relevant memory experts through MMER according to the current hidden state, after which the corresponding suffix $n$-gram contexts are hashed for expert-specific memories lookup. The retrieved memories are subsequently fused into the Transformer backbone through context-aware gated refinement. % Finally, we formulate the training objective of MASC.
As depicted in Fig.~\ref{fig:MASC}, the proposed MASC framework introduces a Mixture-of-Memory-Experts Router (MMER) built on an Engram-style parameterized memory backbone. This architecture performs sparse, hidden-state-dependent expert selection to adaptively retrieve and fuse multi-order local contextual patterns into the Transformer.

% The framework consists of a multi-order memory module with four experts corresponding to \(2\)-gram, \(3\)-gram, \(4\)-gram, and \(5\)-gram patterns, together with a Mixture-of-Memory-Experts Router (MMER) for sparse expert selection. At each coding step, MMER takes the current hidden state as input and activates a sparse subset of relevant memory experts. The selected memory representations are then weighted, fused with the hidden state through context-aware gating, and further refined by a lightweight convolutional module before being injected into the Transformer backbone.

% \subsection{Mixture-of-Memory-Experts Routing}
\subsection{Engram-Style Memory Backbone}
\label{sec:engram}
The Engram-style hashed memory backbone~\cite{Engram} pairs a Transformer for global dependencies with an external hashed memory module for recurring local contextual patterns. %Specifically, as in Eq. \eqref{eq:backbone_memory}, the memory backbone maintains order-specific hashed memories based on short local contexts. %, which can be expressed as
% \begin{equation}
% \mathcal{M}_{\mathrm{Engram}}
% =
% \left\{\mathcal{M}^{(n)}\right\}_{n=2}^{5},
% \end{equation}
% where $\mathcal{M}^{(n)}$ denotes the hashed memory associated with the $n$-gram order.  %The $2$-gram memory captures immediate short-range collocations, while the $3$-gram memory provides slightly longer local contextual patterns.
Given an input token sequence 
$\textbf{t}_{1:N_t}=(t_1,\ldots,t_i,\ldots,t_{N_t})$, 
the suffix context of order $n$ ending at position $i\geq n$ is defined as
\begin{equation}
g_i^{(n)}=(t_{i-n+1},\ldots,t_i),
\qquad n\in\{2,\cdots,5\}.
\end{equation}
To avoid parameter explosion, each suffix context is mapped into a compact set of memory indices through deterministic hashing~\cite{hash}. For the $m$-th hash head of the $n$-gram memory, where $m\in {1, \cdots, M}$, we compute
\begin{equation}
z_i^{(n,m)}=\phi^{(n,m)}\!\left(g_i^{(n)}\right),
\qquad
\textbf{e}_{i}^{(n,m)}=\mathbf{E}^{(n,m)}[z_i^{(n,m)}],
\label{eq:engram_hash_lookup}
\end{equation}
where $\phi^{(n,m)}(\cdot)$ is a hash function and $\mathbf{E}^{(n,m)}$ denotes the corresponding embedding table. 
The final memory representation $\textbf{e}_{i}$ is obtained by concatenating the hashed embeddings $\textbf{e}_{i}^{(n,m)}$, which will be formally given later in Eq. \eqref{eq:masc_routed_memory}.
% :
% \begin{equation}
% \textbf{e}_{i}
% =
% \mathop{\|}_{n\in\{2,3,4,5\}}
% \mathop{\|}_{m=1}^{M}
% \mathbf{e}_{i}^{(n,m)}.
% \label{eq:engram_fixed_aggregation}
% \end{equation}
% where $\|$ denotes the concatenation operation, $M$ is the number of hash heads, and $\textbf{e}_{i}$ denotes the memory representation retrieved from the $n$-gram hashed memories for the contexts ending at token $i$.

During memory fusion, the Transformer hidden state $\textbf{h}_i^{(l)}$ at token $i$ in the $l$-th Transformer layer serves as the query to project the retrieved memory $\textbf{e}_i$ into the key and value spaces:
\begin{equation}
\textbf{q}_i=\textbf{h}_i^{(l)},
\qquad
\textbf{k}_i=\mathbf{W}_K\textbf{e}_i,
\qquad
\textbf{v}_i=\mathbf{W}_V\textbf{e}_i,
\end{equation}
where $\mathbf{W}_K$ and $\mathbf{W}_V$ are learnable projection matrices. The compatibility between the current hidden state and the retrieved memory is measured by
\begin{equation}
\beta_i
=
\sigma\!\left(
\frac{
\mathrm{Norm}(\textbf{q}_i)^{\top}
\mathrm{Norm}(\textbf{k}_i)
}{
\sqrt{d}
}
\right),
\label{eq:engram_gate}
\end{equation}
where $\sigma(\cdot)$ denotes the sigmoid function, $\mathrm{Norm}(\cdot)$ denotes RMS normalization, and $d$ denotes the feature dimension. The gated memory output is then given by
\begin{equation}
\tilde{\textbf{v}}_i=\beta_i\,\textbf{v}_i.
\end{equation}

Collecting the gated memory values of all tokens yields
\begin{equation}
\tilde{\mathbf{V}}^{(l)}
=
\left[
\tilde{\textbf{v}}_1,\tilde{\textbf{v}}_2,\ldots,\tilde{\textbf{v}}_{N_t}
\right].
\end{equation}

To further refine the gated memory sequence and introduce a local nonlinear transformation, the Engram-style backbone applies a lightweight depthwise causal convolution:
\begin{equation}
\mathbf{Y}^{(l)}
=
\mathrm{SiLU}\!\left(
\mathrm{Conv1D}\!\left(
\mathrm{Norm}\!\left(\tilde{\mathbf{V}}^{(l)}\right)
\right)
\right)
+
\tilde{\mathbf{V}}^{(l)},
\label{eq:engram_conv_refine}
\end{equation}
where $\mathrm{Conv1D}(\cdot)$ denotes a lightweight depthwise causal one-dimensional convolution, and $\mathrm{SiLU}(\cdot)$ denotes the Sigmoid Linear Unit activation function.

Finally, the refined memory output is injected into the Transformer backbone through a residual connection:
\begin{equation}
\mathbf{H}^{(l)} \leftarrow \mathbf{H}^{(l)} + \mathbf{Y}^{(l)}.
\label{eq:engram_residual}
\end{equation}

% Although the Engram-style backbone provides an efficient mechanism for retrieving recurring local patterns, its original design mainly considers $2$-gram and $3$-gram memories and relies on a fixed aggregation of the retrieved memory representations. Such a design may be insufficient for source coding, where next-token prediction can depend on contextual patterns with different granularities. To address this limitation, MASC extends the memory orders from $2/3$-gram to $2/3/4/5$-gram and further introduces the proposed MMER mechanism, which adaptively selects the most relevant memory orders according to the current hidden state.

\subsection{MMER-Based Sparse Expert Routing}

% The local semantic pattern most relevant to modeling the next-token distribution may vary across the current token. In some cases, the next-token prediction is primarily determined by short-range collocations and can be sufficiently characterized by lower-order patterns, whereas in others it may depend on longer local contexts that require higher-order memories. As a result, a fixed multi-order memory aggregation strategy may not be sufficiently adaptive to the token-specific contextual requirements. To address this issue, MASC employs an MMER to adaptively select the most relevant memory experts according to the current hidden state.
Since the most informative contextual granularity is token-dependent, MASC introduces an MMER to perform sparse and hidden-state-dependent expert selection. Given the hidden state $\textbf{h}_i^{(l)}$, MMER computes routing logits over the four experts associated with $n$-gram memories as
\begin{equation}
\textbf{r}_i
=
\mathbf{W}_r\,\mathrm{Norm}\!\left(\textbf{h}_i^{(l)}\right)
\in\mathbb{R}^{4},
\end{equation}
where $\mathbf{W}_r$ is a learnable projection matrix. The routing probabilities are then obtained by
\begin{equation}
\boldsymbol{\pi}_i=\mathrm{softmax}(\textbf{r}_i).
\end{equation}
To enforce sparse activation, only the top-$k$ experts are retained:
\begin{equation}
\mathcal{A}_i=\mathrm{TopK}(\boldsymbol{\pi}_i,k)\subseteq \{2,3,4,5\},
\qquad |\mathcal{A}_i|=k.
\end{equation}

Let $\{\alpha_i^{(n)}\}_{n\in\mathcal{A}_i}$ denote the normalized routing weights over the selected experts, given by
\begin{equation}
\alpha_i^{(n)}
=
\frac{\pi_i^{(n)}}{\sum\nolimits_{j\in\mathcal{A}_i}\pi_i^{(j)}},
\qquad n\in\mathcal{A}_i,
\label{eq:mmer_routing_weight}
\end{equation} 
where $\pi_i^{(n)}$ denotes the routing probability assigned to the $n$-gram expert.

% In this way, MMER adaptively identifies the most relevant contextual granularities while avoiding the unnecessary activation of irrelevant  memory experts at every step. 

% \subsection{Expert-Specific Memory Lookup}
\subsection{MMER-Conditioned Memory Retrieval}
% Conditioned on the selected expert subset $\mathcal{A}_i$, MASC retrieves memory representations only from the activated memory experts. For each selected order $n\in\mathcal{A}_i$, the corresponding suffix context is mapped into hashed memory embeddings following the Engram-style lookup mechanism in \eqref{eq:engram_hash_lookup}. The order-specific memory representation is then obtained by concatenating the embeddings over all hash heads:
% \begin{equation}
% \textbf{e}_{i}^{(n)}
% =
% \mathop{\|}_{m=1}^{M}
% \textbf{e}_{i}^{(n,m)},
% \qquad n\in\mathcal{A}_i,
% \label{eq:masc_order_memory}
% \end{equation}
% where $\|$ denotes the concatenation operation, $M$ is the number of hash heads, and $\textbf{e}_{i}^{(n)}$ denotes the memory representation retrieved from the $n$-gram hashed memory for the context ending at token $i$.

Different from the original Engram backbone, which fixedly aggregates the $2$-gram and $3$-gram memories, MASC performs routing-weighted aggregation over the selected experts:
\begin{equation}
\textbf{e}_i
=
\sum_{n\in\mathcal{A}_i}
\alpha_i^{(n)}\,\textbf{e}_{i}^{(n)}.
\label{eq:masc_routed_memory}
\end{equation}
Here, the memory representation $\textbf{e}_{i}^{(n)}$ retrieved from the $n$-gram hashed memory is obtained by concatenating the embeddings from all hash heads: 
\begin{equation}
\textbf{e}_{i}^{(n)}=\mathop{\|}_{m=1}^{M}\textbf{e}_{i}^{(n,m)}, n\in\mathcal{A}_i,
\end{equation}
where $\|$ denotes the concatenation operation and $\alpha_i^{(n)}$ is the normalized routing weight assigned by MMER, as defined in \eqref{eq:mmer_routing_weight}. 
The resulting $\textbf{e}_i$ is then injected into the backbone described in Section \ref{sec:engram}. 

\subsection{Training Objective}

MASC is optimized with a next-token prediction objective, while an auxiliary routing loss is introduced to regularize expert utilization. Given an input token sequence $\textbf{t}_{1:N_t}$, the model is trained via the standard Cross-Entropy (CE) loss:
\begin{equation}
\mathcal{L}_{\mathrm{CE}}
=
-\frac{1}{|\Omega|}
\sum_{i\in\Omega}
\log \tilde{p}_{\mathrm{MASC}}(t_{i+1} \mid \textbf{t}_{1:i}),
\end{equation}
where $\Omega$ denotes the set of valid prediction positions excluding padded tokens, and $\tilde{p}_{\mathrm{MASC}}(\cdot)$ denotes the next-token distribution produced by the MASC source model.

To avoid degenerate routing behavior in MMER, an auxiliary load-balancing loss is introduced to encourage uniform expert utilization. We first define the expert importance as the average soft routing mass:
\begin{equation}
\mathrm{Imp}_n
=
\frac{1}{N_t}
\sum_{i=1}^{N_t}
\pi_i^{(n)},
\qquad n\in\{2,3,4,5\},
\end{equation}
and the expert load as the average hard selection frequency:
\begin{equation}
\mathrm{Load}_n
=
\frac{1}{kN_t}
\sum_{i=1}^{N_t}
\mathbb{I}\!\left(n\in\mathcal{A}_i\right),
\qquad n\in\{2,3,4,5\},
\end{equation}
where $\mathbb{I}(\cdot)$ is the indicator function. Based on these quantities, the auxiliary routing loss is defined as
\begin{equation}
\mathcal{L}_{\mathrm{aux}}
=
4\sum_{n\in\{2,3,4,5\}}
\mathrm{Imp}_n\,\mathrm{Load}_n.
\end{equation}

The overall training objective is then given by
\begin{equation}
\mathcal{L}
=
\mathcal{L}_{\mathrm{CE}}
+
\lambda_{\mathrm{aux}}\mathcal{L}_{\mathrm{aux}},
\label{eq:total_loss}
\end{equation}
where $\lambda_{\mathrm{aux}}$ is a balancing coefficient. %In this way, the CE term drives next-token prediction, while the auxiliary term regularizes expert utilization and stabilizes sparse routing during training.

\section{Simulation Settings and Results}\label{sec4_Experiment}

\subsection{Simulation Settings}

\begin{figure}[!t]
\centering
\includegraphics[width=\linewidth]{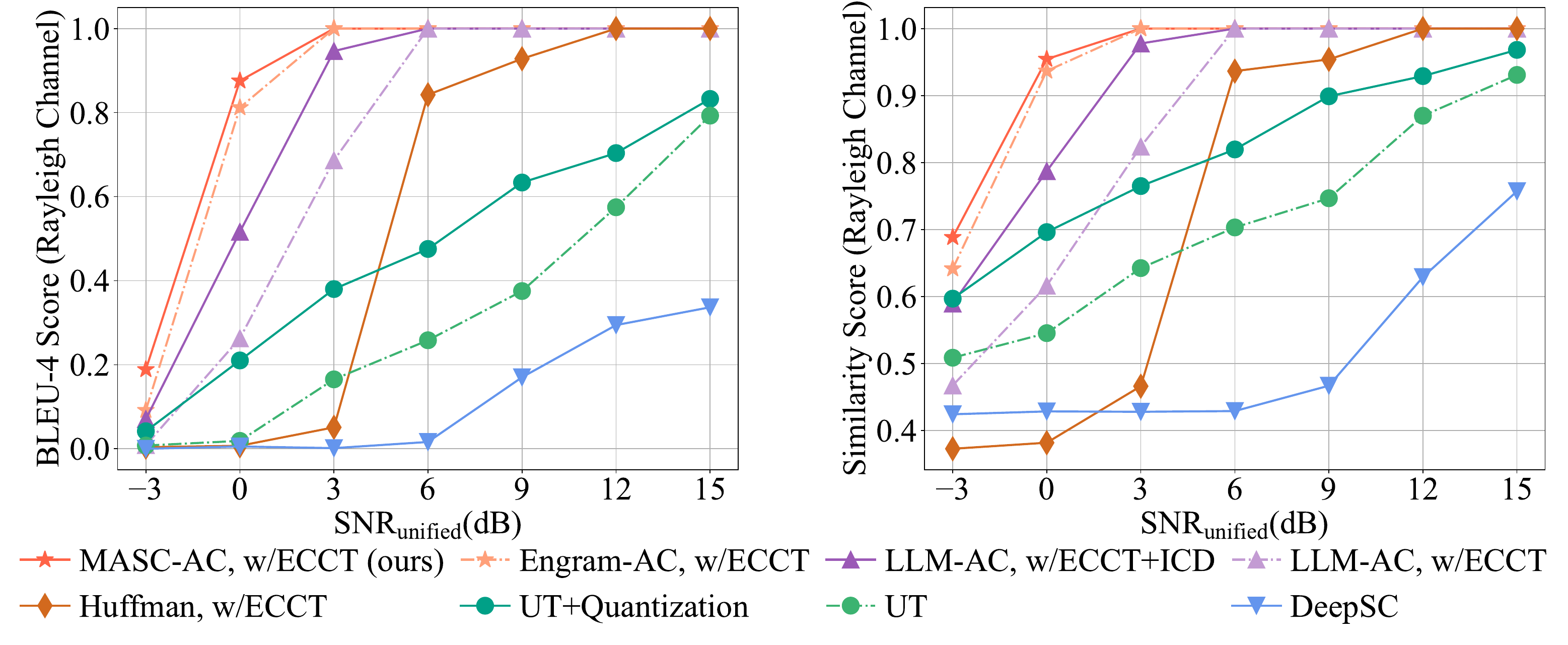}
\vspace{-.7cm}
\caption{Overall system-level performance under Rayleigh channels.}
\label{fig:overall_performance_rayleigh}
\end{figure}

\begin{figure}[!t]
\centering
\includegraphics[width=\linewidth]{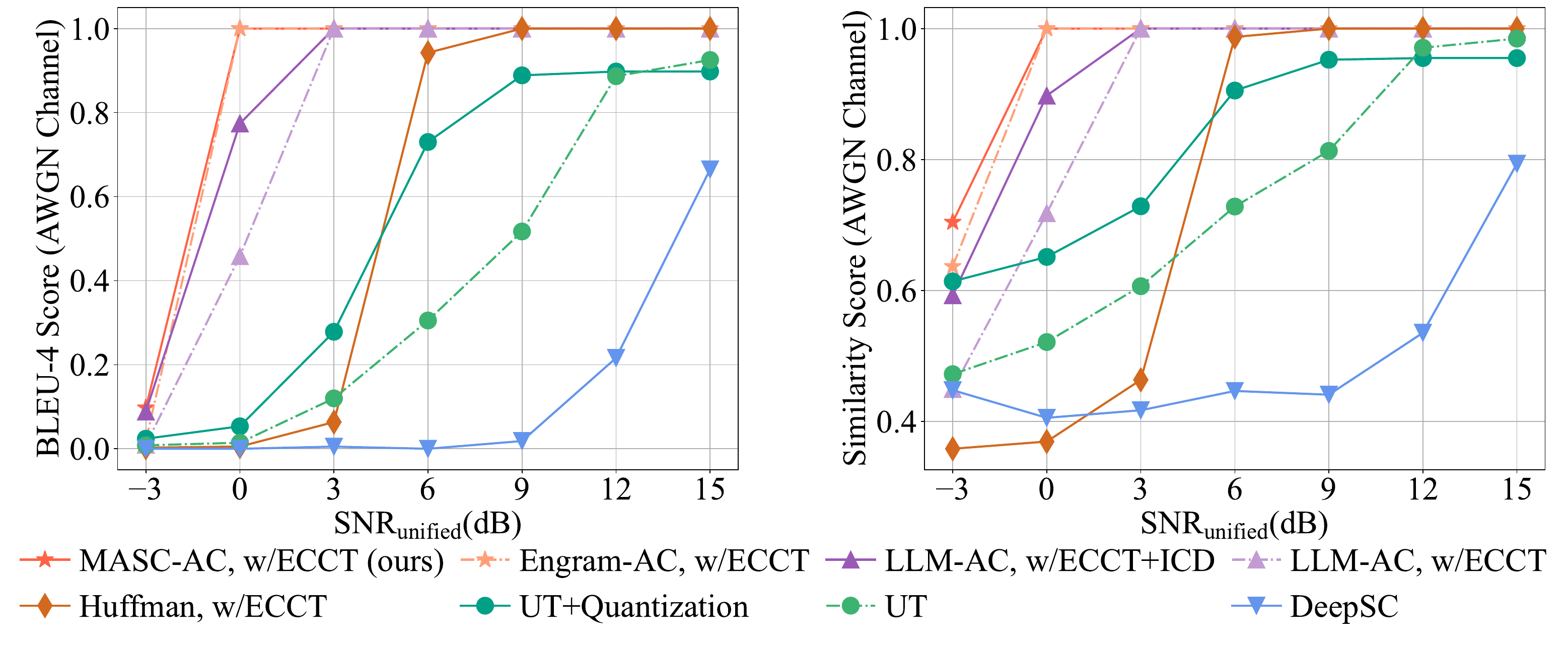}
\vspace{-.7cm}
\caption{Overall system-level performance under AWGN channels.}
\label{fig:overall_performance_awgn}
\end{figure}

We evaluate the proposed MASC scheme against conventional SSCC pipelines and representative JSCC-based semantic communication schemes over both Rayleigh fading and AWGN channels using the English portion of the Europarl corpus~\cite{dataset}. The raw text is preprocessed via standard sentence splitting and deduplication, and randomly partitioned into training, validation, and test sets with an $80\%/10\%/10\%$ ratio.
% To ensure a fair and reproducible comparison, all methods are tested on the same text source and are assessed using widely adopted Natural Language Processing (NLP)-oriented quality metrics.

% \textbf{Dataset and source coding.}
% The adopted dataset is the English portion of the Europarl corpus~\cite{dataset}. We preprocess the raw text by removing XML tags and speaker identifiers, splitting long lines into sentence-level samples at punctuation marks, and removing duplicate sentences. The resulting samples are then randomly shuffled with a fixed seed and partitioned into training, validation, and test sets with a ratio of $80\%/10\%/10\%$. For the conventional SSCC baselines, including LLM-AC with ECCT and LLM-AC with ECCT and ICD, we adopt a standard GPT-$2$-based source model. In contrast, Engram-AC with ECCT and the proposed MASC employ their own trained memory-augmented source models, with the latter further integrating a PCM equipped with an MMER. The arithmetic coder is configured with a precision of $31$ bits, which provides a practical trade-off between numerical stability and coding accuracy.

\textbf{Source and Channel Coding Configurations.}
For the conventional SSCC baselines, we employ a standard GPT-$2$-based source model (124.44M parameters), whereas Engram-AC (149.76M parameters) and MASC (117.98M parameters) utilize their respective trained memory-augmented source models. The arithmetic coder is configured with a $31$-bit precision. For channel coding, all SSCC pipelines employ an LDPC$(49,24)$ code (code rate $\approx 0.5$), complemented by an ECCT decoder at the receiver side to mitigate noise. 
% The ECCT architecture and training hyperparameters, as well as those of the JSCC baselines, are summarized in Table~\ref{tab2:settings}.

\textbf{Baselines.}
We include the following representative baselines:
\begin{itemize}
    \item \textbf{Engram-AC with ECCT}~\cite{Engram}, which enhances LLM-based AC with a classical Engram memory.
    \item \textbf{LLM-AC with ECCT and ICD}~\cite{ICD}, which exploits receiver-side contextual information and candidate-based reconstruction.
    \item \textbf{LLM-AC with ECCT}~\cite{SSCC1}, which uses GPT-$2$-based AC.
    \item \textbf{Huffman with ECCT}, which replaces LLM-AC with Huffman coding, while keeping the rest of the SSCC pipeline unchanged.
    \item \textbf{DeepSC}~\cite{DeepSC} and \textbf{UT}~\cite{UT}, which serve as representative JSCC baselines for comparison.
    \item \textbf{UT with quantization}~\cite{UT_quanti}, where the latent representation is mapped to a fixed $30$-bit stream.
\end{itemize}

\textbf{Evaluation metrics.}
We evaluate reconstruction quality using Bilingual Evaluation Understudy (BLEU) and a Bidirectional Encoder Representations from Transformers (BERT)-based semantic similarity metric, which measure lexical fidelity and semantic preservation, respectively. To fairly compare heterogeneous pipelines that may transmit different physical-layer payload lengths $N$ for the same source content, we follow~\cite{SSCC1} and adopt a unified SNR under a fixed total transmission energy budget.

\subsection{Overall System-Level Performance Comparison}

Fig.~\ref{fig:overall_performance_rayleigh} and Fig.~\ref{fig:overall_performance_awgn} compare the proposed method with multiple baselines under Rayleigh fading and AWGN channels, respectively, using BLEU-4 and cosine similarity. It is observed that MASC consistently outperforms the conventional SSCC baselines, including Huffman with ECCT and LLM-AC with ECCT, as well as the representative JSCC schemes. MASC also achieves uniformly better performance than the Engram-based scheme, confirming the effectiveness of introducing MMER-based sparse expert selection on top of the shared memory-augmented source modeling framework. The gain is particularly pronounced in the low-SNR regime, where improved conditional probability modeling yields shorter codewords and thus mitigates the susceptibility of arithmetic decoding to residual channel errors. Furthermore, MASC reliably exhibits better performance than ICD. It is worth emphasizing that ICD operates under a stronger assumption of perfectly reliable contextual information at the receiver, whereas MASC achieves robustness enhancement through transceiver-level memory-augmented source modeling in a self-contained manner.

% \subsection{Comparison with Larger LLM Backbones}
% Required packages:
% \usepackage{booktabs}
% \usepackage{multirow}
% \usepackage{array}

% \begin{table*}[t]
% \centering
% \caption{Performance comparison under different numbers of activated MMER experts.}
% \label{tab:moe_expert_ablation}

% \renewcommand{\arraystretch}{1.15}
% \setlength{\tabcolsep}{8pt}

% \begin{tabular}{c|cccc}
% \toprule
% \multirow{2}{*}{SNR (dB)} 
% & \multicolumn{4}{c}{BLEU4} \\
% \cmidrule(lr){2-5}
% & 1 Expert & 2 Experts & 3 Experts & 4 Experts \\
% \midrule
% -3 
% & 0.1106 & \textbf{0.1887} & 0.0820 & 0.1234 \\
% 0  
% & 0.7534 & \textbf{0.8756} & 0.8588 & 0.7012 \\
% \bottomrule
% \end{tabular}
% \hspace{1.2cm}
% \begin{tabular}{c|cccc}
% \toprule
% \multirow{2}{*}{SNR (dB)} 
% & \multicolumn{4}{c}{Cosine Similarity} \\
% \cmidrule(lr){2-5}
% & 1 Expert & 2 Experts & 3 Experts & 4 Experts \\
% \midrule
% -3 
% & 0.6339 & \textbf{0.6884} & 0.6581 & 0.6459 \\
% 0  
% & 0.9266 & 0.9544 & \textbf{0.9641} & 0.9137 \\
% \bottomrule
% \end{tabular}

% \vspace{2mm}
% \begin{minipage}{0.95\linewidth}
% \footnotesize
% \textit{Note:} Bold values indicate the best result for each SNR.
% The configuration with 2 experts is the default setting.
% \end{minipage}

% \end{table*}

\begin{table}[!t]
\centering
\caption{Performance comparison with different numbers of activated MMER experts over Rayleigh channels.}
\label{tab:moe_expert_ablation}

\renewcommand{\arraystretch}{1.1}
\setlength{\tabcolsep}{3.5pt}
\setlength{\textfloatsep}{6pt}
\setlength{\floatsep}{4pt}
\setlength{\intextsep}{6pt}

\begin{tabular}{c|cccc|cccc}
\toprule
\multirow{2}{*}{\makecell{SNR\\(dB)}}
& \multicolumn{4}{c|}{BLEU4}
& \multicolumn{4}{c}{Cosine Similarity} \\
\cmidrule(lr){2-5}\cmidrule(l){6-9}
% & 1 Exp. & 2 Exp. & 3 Exp. & 4 Exp.
% & 1 Exp. & 2 Exp. & 3 Exp. & 4 Exp. \\
& $k=1$ & $k=2$ & $k=3$ & $k=4$
& $k=1$ & $k=2$ & $k=3$ & $k=4$ \\
\midrule
$-3$
& $0.111$ & $\mathbf{0.189}$ & $0.082$ & $0.123$
& $0.634$ & $\mathbf{0.688}$ & $0.658$ & $0.646$ \\
$0$
& $0.753$ & $\mathbf{0.876}$ & $0.859$ & $0.701$
& $0.927$ & $0.954$ & $\mathbf{0.964}$ & $0.914$ \\
\bottomrule
\end{tabular}

\vspace{1mm}

\begin{minipage}{0.95\linewidth}
\footnotesize
\textit{Note:} Bold values indicate the best result for each SNR.
The configuration with 2 experts is the default setting.
\end{minipage}

\end{table}

\subsection{Impact of the Number of Activated MMER Experts}
\label{sec:expert_ablation}

Table~\ref{tab:moe_expert_ablation} investigates the effect of different numbers of activated MMER experts. The two-expert configuration generally provides a favorable performance trade-off, indicating that moderate expert activation can effectively balance contextual diversity and sparse routing efficiency.

\section{Conclusions}\label{sec5_Conclusions}
In this letter, we have presented MASC, a memory-augmented source coding scheme to mitigate the cliff effect in SSCC-based transmission. 
% By endowing both the transmitter and the receiver with a shared PCM, MASC has incorporated recurring local semantic patterns directly into source model through multi-order $n$-gram memories. To make such memory expressive yet efficient, MMER has performed sparse expert routing over $2$-gram, $3$-gram, $4$-gram, and $5$-gram memory experts during autoregressive source modeling.
By integrating the shared PCM with the MMER, MASC adaptively internalizes multi-order contextual patterns to refine source probability estimation.
Simulation results over Rayleigh fading and AWGN channels demonstrate its superior performance over conventional SSCC, JSCC, and receiver-side correction methods. In future work, we will extend this paradigm toward richer memory representations and broader multimodal transmission scenarios.
\bibliographystyle{IEEEtran}
\bibliography{reference}
\end{document}